\documentclass[conference,10pt]{IEEEtran}
\IEEEoverridecommandlockouts
\usepackage{cite}
\usepackage{amsmath,amssymb,amsfonts}
\usepackage{algorithmic}
\usepackage{graphicx}
\usepackage{textcomp}
\usepackage{xcolor}
\def\BibTeX{{\rm B\kern-.05em{\sc i\kern-.025em b}\kern-.08em
    T\kern-.1667em\lower.7ex\hbox{E}\kern-.125emX}}

\usepackage{amsmath,graphicx}

\usepackage{microtype}
\usepackage[english]{babel}   
\usepackage[ansinew]{inputenc}
\usepackage[T1]{fontenc}      
\usepackage[sort&compress,numbers]{natbib}	
\usepackage[colorlinks=true,pdfborder={0 0 0}]{hyperref}
\usepackage{units}
\usepackage{import}
\usepackage{balance}

\usepackage[acronym,nohypertypes={acronym,notation}]{glossaries} 
\usepackage{url}

\usepackage{trfsigns} 
\usepackage{rotating} 
\usepackage{xspace} 

\usepackage{amsfonts,amssymb}
\usepackage{array}
\usepackage{bm} 
\usepackage{bbm} 
\usepackage{siunitx} 

\usepackage{subfig}
\usepackage{xcolor}
\usepackage{tikz}
\usetikzlibrary{calc}
\usetikzlibrary{tikzmark}

\usepackage{longtable}
\usepackage{tabularx}
\usepackage{balance}

\usepackage{tikz}
\usepackage{pgfplots}
\usepackage{amssymb}
\usetikzlibrary{positioning,chains,calc,shapes.geometric, shadows, shapes.misc, fit, arrows, shapes.callouts}

\usepackage{tabularx, etoolbox, booktabs}
\usepackage{multirow} 
\usepackage{bbding}
\usepackage{microtype}
\usepackage{pifont}

\usepackage{trfsigns}
\newcommand{\GG}{$\Gamma(t,f_{P}(t),f_{D})$}
\newcommand{\fD}{f_D}

\DeclareMathOperator{\argmax}{argmax}

\newacronym{ML}{ML}{maximum likelihood}
\newacronym{MAP}{MAP}{maximum a-posteriori}

\newacronym{DNN}{DNN}{deep neural network}
\newacronym{CDR}{CDR}{coherent-to-diffuse power ratio}
\newacronym{DoA}{DoA}{direction-of-arrival}
\newacronym{TDoA}{TDoA}{time difference of arrival}
\newacronym{MSE}{MSE}{mean-squared error}
\newacronym{MLP}{MLP}{multilayer perceptron}
\newacronym{CRNN}{CRNN}{convolutional recurrent neural network}
\newacronym{RNN}{RNN}{recurrent neural network}
\newacronym[plural=GPs,firstplural=Gaussian processes (GPs)]{GP}{GP}{Gaussian process}
\newacronym{GRU}{GRU}{gated recurrent unit}
\newacronym{RIR}{RIR}{room impulse response}
\newacronym{AWGN}{AWGN}{additive white Gaussian noise}
\newacronym{SNR}{SNR}{signal-to-noise ratio}
\newacronym{STFT}{STFT}{short-time Fourier transform}
\newacronym[plural=PSDs,firstplural=power spectral densities~(PSDs)]{PSD}{PSD}{power sprectral density}
\newacronym{AE}{AE}{absolute error}
\newacronym{MAE}{MAE}{mean-absolute error}
\newacronym{PE}{PE}{position error}
\newacronym{MPE}{MPE}{mean position error}
\newacronym{WASN}{WASN}{wireless acoustic sensor network}
\newacronym{OoR}{OoR}{out-of-range}
\newacronym{ASR}{ASR}{automatic speech recognition}
\newacronym{TDNN}{TDNN}{time delay neural network}
\newacronym{CNN}{CNN}{convolutional neural network}
\newacronym{WLS}{WLS}{weighted least squares}
\newacronym{LS}{LS}{least squares}
\newacronym{RANSAC}{RANSAC}{random sample consensus}

\newacronym{GARDE}{GARDE}{Geometry cAlibration fRom Distance Estimates}
\newacronym{MDS}{MDS}{Multi Dimensional Scaling}

\newacronym{GMM}{GMM}{gaussian mixture model}
\newacronym{HMM}{HMM}{hidden Markov model}

\newacronym{CWLS}{CWLS}{constrained weighted least squares}
\newacronym{CRLB}{CRLB}{Cramer-Rao lower bound}
\newacronym{RMSE}{RMSE}{Root Mean Square Error}

\newacronym{ITU}{ITU}{International Telecommunication Union}
\newacronym{SDR}{SDR}{software defined radio}
\newacronym{SAD}{SAD}{speech activity detection}
\newacronym{ROC}{ROC}{receiver operating characteristic}
\newacronym{AUC}{AUC}{area under the curve}
\newacronym{EER}{EER}{equal error rate}
\newacronym{AGC}{AGC}{Automatic Gain Control}
\newacronym{STOI}{STOI}{Short-time Objective Intelligibility}
\newacronym{CSBE}{CSBE}{combined sub-band energy}
\newacronym{SSB}{SSB}{single sideband}
\newacronym{HF}{HF}{high frequency}
\newacronym{RT}{RT}{realtime}

\newacronym{SHC}{SHC}{Spectral Harmonics Distortion}
\newacronym{FFT}{FFT}{Fast Fourier Transformation}

\newacronym{LSB}{LSB}{Lower Side Band}
\newacronym{USB}{USB}{Upper Side Band}

\usepackage{tikz}
\usepackage{standalone}
\usepackage{pgfplots}

\usetikzlibrary{positioning,chains,calc,shapes.geometric, shadows, shapes.misc, fit, arrows, shapes.callouts}
\usepgfplotslibrary{groupplots}
\pgfplotsset{compat=newest}
\usetikzlibrary{backgrounds}

\definecolor{yellow}{RGB}{255, 198, 0}
\definecolor{orange}{RGB}{255, 130, 0}
\definecolor{blue}{RGB}{0, 32, 91}
\definecolor{red}{RGB}{198, 53, 39}
\definecolor{magenta}{RGB}{138, 27, 97}
\definecolor{lightblue}{RGB}{0, 159, 223}
\definecolor{green}{RGB}{0, 155,119}
\definecolor{lightgreen}{RGB}{132, 189,0}
\definecolor{greenyellow}{RGB}{208, 223,0}

\begin{document}

\title{Open Range Pitch Tracking for Carrier Frequency Difference Estimation from HF Transmitted Speech \thanks{This work has been partially funded by the Plath GmbH, Hamburg.}
}

\author{Joerg Schmalenstroeer, Jens Heitkaemper, Joerg Ullmann, Reinhold Haeb-Umbach\\
\textit{Department of Communications Engineering, Paderborn University, Germany}\\	
\footnotesize Email: \texttt{\{schmalen, jensheit, ullmann, haeb\}@nt.uni-paderborn.de}
}

\maketitle

\begin{abstract}
	In this paper we investigate the task of detecting carrier frequency differences from demodulated single sideband signals by examining the pitch contours of the received baseband speech signal in the short-time spectral domain. From the  detected pitch frequency trajectory and its harmonics a carrier frequency difference, which is caused by demodulating the radio signal with the wrong carrier frequency, can be deduced. A computationally efficient realization in the power cepstral domain is presented. The core component, i.e., the pitch tracking algorithm, is shown to perform comparably to a state of the art algorithm. The full carrier frequency difference estimation system is tested on recordings of real transmissions over HF links.  A comparison with an existing approach shows improved estimation accuracy, both on short and longer  speech utterances.
	
	%
\end{abstract}
%


\section{Introduction}
\begin{figure}[b]
	\centering
%
%
\begin{tikzpicture}

\begin{axis}[%
width=2.7in,
height=1.2in,
at={(0in,0in)},
scale only axis,
axis on top,
xmin=-0.008,
xmax=2.52,
xlabel={Time [s]},
ymin=-0.0009765625,
ymax=2,
ylabel={Frequency [kHz]},
legend style={legend cell align=left,align=left,draw=white!15!black}
]
\addplot [forget plot] graphics [xmin=-0.008,xmax=2.52,ymin=-0.0009765625,ymax=3.9990234375] {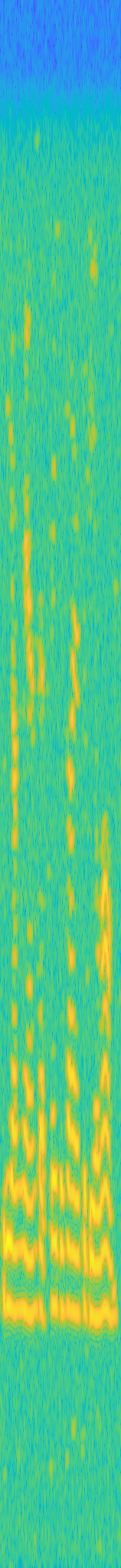};
\addplot [color=black,solid,forget plot]
  table[row sep=crcr]{%
0.064	0.593359375\\
0.08	0.577734375\\
0.096	0.655859375\\
0.112	0.65390625\\
0.128	0.655859375\\
0.144	0.663671875\\
0.16	0.663671875\\
0.176	0.665625\\
0.192	0.66171875\\
0.208	0.659765625\\
0.224	0.659765625\\
0.24	0.6578125\\
0.256	0.655859375\\
0.272	0.651953125\\
0.288	0.651953125\\
0.304	0.65\\
0.32	0.651953125\\
0.336	0.651953125\\
0.352	0.651953125\\
0.368	0.651953125\\
0.384	0.65390625\\
0.4	0.655859375\\
0.416	0.655859375\\
0.432	0.655859375\\
0.448	0.65390625\\
0.464	0.655859375\\
0.48	0.655859375\\
0.496	0.65390625\\
0.512	0.651953125\\
0.528	0.65\\
0.544	0.648046875\\
0.56	0.644140625\\
0.576	0.6421875\\
0.592	0.640234375\\
0.608	0.640234375\\
0.624	0.640234375\\
0.64	0.640234375\\
0.656	0.640234375\\
0.672	0.6421875\\
0.688	0.644140625\\
0.704	0.64609375\\
};
\addplot [color=black,solid,forget plot]
  table[row sep=crcr]{%
0.064	0.68515625\\
0.08	0.65390625\\
0.096	0.81015625\\
0.112	0.80625\\
0.128	0.81015625\\
0.144	0.82578125\\
0.16	0.82578125\\
0.176	0.8296875\\
0.192	0.821875\\
0.208	0.81796875\\
0.224	0.81796875\\
0.24	0.8140625\\
0.256	0.81015625\\
0.272	0.80234375\\
0.288	0.80234375\\
0.304	0.7984375\\
0.32	0.80234375\\
0.336	0.80234375\\
0.352	0.80234375\\
0.368	0.80234375\\
0.384	0.80625\\
0.4	0.81015625\\
0.416	0.81015625\\
0.432	0.81015625\\
0.448	0.80625\\
0.464	0.81015625\\
0.48	0.81015625\\
0.496	0.80625\\
0.512	0.80234375\\
0.528	0.7984375\\
0.544	0.79453125\\
0.56	0.78671875\\
0.576	0.7828125\\
0.592	0.77890625\\
0.608	0.77890625\\
0.624	0.77890625\\
0.64	0.77890625\\
0.656	0.77890625\\
0.672	0.7828125\\
0.688	0.78671875\\
0.704	0.790625\\
};
\addplot [color=black,solid,forget plot]
  table[row sep=crcr]{%
0.064	0.776953125\\
0.08	0.730078125\\
0.096	0.964453125\\
0.112	0.95859375\\
0.128	0.964453125\\
0.144	0.987890625\\
0.16	0.987890625\\
0.176	0.99375\\
0.192	0.98203125\\
0.208	0.976171875\\
0.224	0.976171875\\
0.24	0.9703125\\
0.256	0.964453125\\
0.272	0.952734375\\
0.288	0.952734375\\
0.304	0.946875\\
0.32	0.952734375\\
0.336	0.952734375\\
0.352	0.952734375\\
0.368	0.952734375\\
0.384	0.95859375\\
0.4	0.964453125\\
0.416	0.964453125\\
0.432	0.964453125\\
0.448	0.95859375\\
0.464	0.964453125\\
0.48	0.964453125\\
0.496	0.95859375\\
0.512	0.952734375\\
0.528	0.946875\\
0.544	0.941015625\\
0.56	0.929296875\\
0.576	0.9234375\\
0.592	0.917578125\\
0.608	0.917578125\\
0.624	0.917578125\\
0.64	0.917578125\\
0.656	0.917578125\\
0.672	0.9234375\\
0.688	0.929296875\\
0.704	0.93515625\\
};
\addplot [color=black,solid,forget plot]
  table[row sep=crcr]{%
0.816	0.651953125\\
0.832	0.65390625\\
0.848	0.651953125\\
0.864	0.65\\
0.88	0.644140625\\
0.896	0.624609375\\
0.912	0.624609375\\
0.928	0.624609375\\
};
\addplot [color=black,solid,forget plot]
  table[row sep=crcr]{%
0.816	0.80234375\\
0.832	0.80625\\
0.848	0.80234375\\
0.864	0.7984375\\
0.88	0.78671875\\
0.896	0.74765625\\
0.912	0.74765625\\
0.928	0.74765625\\
};
\addplot [color=black,solid,forget plot]
  table[row sep=crcr]{%
0.816	0.952734375\\
0.832	0.95859375\\
0.848	0.952734375\\
0.864	0.946875\\
0.88	0.929296875\\
0.896	0.870703125\\
0.912	0.870703125\\
0.928	0.870703125\\
};
\addplot [color=black,solid,forget plot]
  table[row sep=crcr]{%
1.056	0.66171875\\
1.072	0.66171875\\
1.088	0.663671875\\
1.104	0.663671875\\
1.12	0.663671875\\
1.136	0.663671875\\
1.152	0.66171875\\
1.168	0.663671875\\
1.184	0.663671875\\
};
\addplot [color=black,solid,forget plot]
  table[row sep=crcr]{%
1.056	0.821875\\
1.072	0.821875\\
1.088	0.82578125\\
1.104	0.82578125\\
1.12	0.82578125\\
1.136	0.82578125\\
1.152	0.821875\\
1.168	0.82578125\\
1.184	0.82578125\\
};
\addplot [color=black,solid,forget plot]
  table[row sep=crcr]{%
1.056	0.98203125\\
1.072	0.98203125\\
1.088	0.987890625\\
1.104	0.987890625\\
1.12	0.987890625\\
1.136	0.987890625\\
1.152	0.98203125\\
1.168	0.987890625\\
1.184	0.987890625\\
};
\addplot [color=black,solid,forget plot]
  table[row sep=crcr]{%
1.248	0.65\\
1.264	0.655859375\\
1.28	0.65390625\\
1.296	0.65\\
1.312	0.65\\
1.328	0.64609375\\
};
\addplot [color=black,solid,forget plot]
  table[row sep=crcr]{%
1.248	0.7984375\\
1.264	0.81015625\\
1.28	0.80625\\
1.296	0.7984375\\
1.312	0.7984375\\
1.328	0.790625\\
};
\addplot [color=black,solid,forget plot]
  table[row sep=crcr]{%
1.248	0.946875\\
1.264	0.964453125\\
1.28	0.95859375\\
1.296	0.946875\\
1.312	0.946875\\
1.328	0.93515625\\
};
\addplot [color=black,solid,forget plot]
  table[row sep=crcr]{%
1.392	0.66171875\\
1.408	0.659765625\\
1.424	0.655859375\\
1.44	0.65390625\\
1.456	0.651953125\\
1.472	0.65\\
1.488	0.65\\
1.504	0.65\\
1.52	0.65\\
1.536	0.648046875\\
1.552	0.648046875\\
1.568	0.64609375\\
1.584	0.644140625\\
1.6	0.644140625\\
1.616	0.644140625\\
1.632	0.644140625\\
1.648	0.6421875\\
};
\addplot [color=black,solid,forget plot]
  table[row sep=crcr]{%
1.392	0.821875\\
1.408	0.81796875\\
1.424	0.81015625\\
1.44	0.80625\\
1.456	0.80234375\\
1.472	0.7984375\\
1.488	0.7984375\\
1.504	0.7984375\\
1.52	0.7984375\\
1.536	0.79453125\\
1.552	0.79453125\\
1.568	0.790625\\
1.584	0.78671875\\
1.6	0.78671875\\
1.616	0.78671875\\
1.632	0.78671875\\
1.648	0.7828125\\
};
\addplot [color=black,solid,forget plot]
  table[row sep=crcr]{%
1.392	0.98203125\\
1.408	0.976171875\\
1.424	0.964453125\\
1.44	0.95859375\\
1.456	0.952734375\\
1.472	0.946875\\
1.488	0.946875\\
1.504	0.946875\\
1.52	0.946875\\
1.536	0.941015625\\
1.552	0.941015625\\
1.568	0.93515625\\
1.584	0.929296875\\
1.6	0.929296875\\
1.616	0.929296875\\
1.632	0.929296875\\
1.648	0.9234375\\
};
\addplot [color=black,solid,forget plot]
  table[row sep=crcr]{%
1.744	0.66171875\\
1.76	0.655859375\\
1.776	0.64609375\\
1.792	0.634375\\
1.808	0.628515625\\
};
\addplot [color=black,solid,forget plot]
  table[row sep=crcr]{%
1.744	0.821875\\
1.76	0.81015625\\
1.776	0.790625\\
1.792	0.7671875\\
1.808	0.75546875\\
};
\addplot [color=black,solid,forget plot]
  table[row sep=crcr]{%
1.744	0.98203125\\
1.76	0.964453125\\
1.776	0.93515625\\
1.792	0.9\\
1.808	0.882421875\\
};
\addplot [color=black,solid,forget plot]
  table[row sep=crcr]{%
1.872	0.634375\\
1.888	0.640234375\\
1.904	0.640234375\\
1.92	0.640234375\\
1.936	0.636328125\\
1.952	0.636328125\\
1.968	0.634375\\
1.984	0.634375\\
2	0.632421875\\
2.016	0.632421875\\
2.032	0.632421875\\
2.048	0.632421875\\
2.064	0.632421875\\
2.08	0.634375\\
2.096	0.636328125\\
2.112	0.64609375\\
2.128	0.648046875\\
2.144	0.65\\
2.16	0.651953125\\
2.176	0.655859375\\
2.192	0.655859375\\
2.208	0.655859375\\
2.224	0.65390625\\
2.24	0.65\\
2.256	0.64609375\\
2.272	0.6421875\\
2.288	0.634375\\
2.304	0.63046875\\
2.32	0.63046875\\
2.336	0.632421875\\
};
\addplot [color=black,solid,forget plot]
  table[row sep=crcr]{%
1.872	0.7671875\\
1.888	0.77890625\\
1.904	0.77890625\\
1.92	0.77890625\\
1.936	0.77109375\\
1.952	0.77109375\\
1.968	0.7671875\\
1.984	0.7671875\\
2	0.76328125\\
2.016	0.76328125\\
2.032	0.76328125\\
2.048	0.76328125\\
2.064	0.76328125\\
2.08	0.7671875\\
2.096	0.77109375\\
2.112	0.790625\\
2.128	0.79453125\\
2.144	0.7984375\\
2.16	0.80234375\\
2.176	0.81015625\\
2.192	0.81015625\\
2.208	0.81015625\\
2.224	0.80625\\
2.24	0.7984375\\
2.256	0.790625\\
2.272	0.7828125\\
2.288	0.7671875\\
2.304	0.759375\\
2.32	0.759375\\
2.336	0.76328125\\
};
\addplot [color=black,solid,forget plot]
  table[row sep=crcr]{%
1.872	0.9\\
1.888	0.917578125\\
1.904	0.917578125\\
1.92	0.917578125\\
1.936	0.905859375\\
1.952	0.905859375\\
1.968	0.9\\
1.984	0.9\\
2	0.894140625\\
2.016	0.894140625\\
2.032	0.894140625\\
2.048	0.894140625\\
2.064	0.894140625\\
2.08	0.9\\
2.096	0.905859375\\
2.112	0.93515625\\
2.128	0.941015625\\
2.144	0.946875\\
2.16	0.952734375\\
2.176	0.964453125\\
2.192	0.964453125\\
2.208	0.964453125\\
2.224	0.95859375\\
2.24	0.946875\\
2.256	0.93515625\\
2.272	0.9234375\\
2.288	0.9\\
2.304	0.88828125\\
2.32	0.88828125\\
2.336	0.894140625\\
};
\addplot [color=black,dashed,forget plot, line width=0.4mm]
  table[row sep=crcr]{%
0	0.5015625\\
2.512	0.5015625\\
};
\node[right, align=left, inner sep=0mm, text=black]
at (axis cs:0.65,0.38,0) {Carrier frequency difference};
\end{axis}
\end{tikzpicture}%
	\caption{Spectrogram of a signal transmitted over HF and demodulated with \SI{500}{Hz} carrier frequency difference. Marked in black are the carrier frequency difference (dashed) and the pitch traces including two harmonics.}
	\label{fig:spec}
\end{figure}
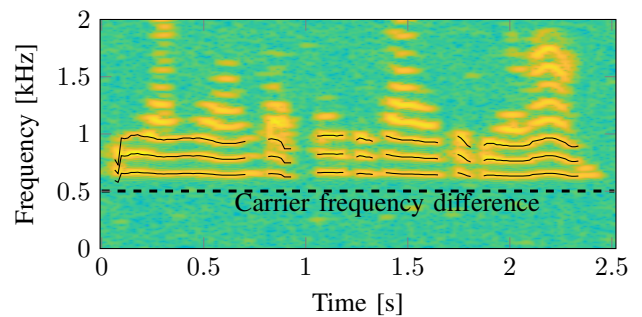

The scenario at hand envisions a  radio station listening on a fixed, pre-selected frequency, and seeking for \gls{SSB} modulated \gls{HF} signals. 
If the receiver selects a different carrier frequency than the transmitter, the demodulated signal contains a frequency shifted version of the original speech signal originating from the carrier frequency difference \cite{Suzuki1994}. This has a  detrimental effect on the intelligibility of the transmitted speech signal \cite{Baskent2007}.  Fig.~\ref{fig:spec} shows the spectrogram of a demodulated signal from a  station operating at a carrier frequency difference of $\SI{500}{Hz}$ compared to the transmitter. 

To improve intelligibility, the carrier frequency difference should be estimated and the signal shifted in frequency to remove the difference. This contribution is concerned with the first task, the determination of the carrier frequency difference from the demodulated speech signal. The second task, the compensation, is rather straightforward and will not be considered here.

A carrier frequency difference can be estimated by investigating the statistical properties of speech, e.g., the modulation symmetry \cite{Clark2013} or the spectral envelope \cite{Ganapathy2013}. 
The contribution \cite{Clark2013} utilizes a third-order modulation spectral analysis that, however, limits the analyzable spectrum  to one-forth of its total width, and \cite{Ganapathy2013} proposes a fundamental harmonic frequency detector that requires relatively long speech segments for reliable estimation in noisy conditions. 
Training based frequency offset estimation has been proposed in \cite{Xing2017}, where GMM-SVMs, i-Vectors and deep neural networks are employed. The main disadvantage here is the requirement of having a representative and large enough data set for training, as \gls{HF} transmissions include a variety of distortions.

In this paper we follow the idea of \cite{Suzuki1994, Ganapathy2013, Xing2017}: By detecting the typical pitch structures in the spectrogram, a possible carrier frequency difference becomes apparent. 
Fundamental frequency estimation, or pitch tracking, has been a research topic  for years with applications in signal enhancement and speaker identification tasks. Various approaches are known from the literature, e.g., RAPT \cite{Talkin2005ARA}, STRAIGHT \cite{Kawahara02}, YIN \cite{YIN02} and YAAPT \cite{Zahorian08}.
Since most  approaches are based on correlation techniques, be it in the time \cite{YIN02} or frequency domain \cite{Kawahara02} or even in both domains \cite{Zahorian08}, comparative studies show only small differences between the algorithms in terms of precision \cite{8081482} as they all depend on similar features. Detecting candidates for periodic signals within the physical range of the vocal cord's oscillation frequencies is usually the first step, which is followed by a post-processing for candidate refinement and subsequent smoothing \cite{Talkin2005ARA, Zahorian08}. Besides time and frequency domain, also cepstral domain estimators have been proposed \cite{Gerkman2010}. 
Clearly, all methods suffer from low \gls{SNR} ratios \cite{8081482} and robustness to distortions is an important aspect. Here, new approaches based on \glspl{DNN} reported promising results \cite{Han2014}.

However, pitch tracking with the purpose of frequency difference estimation poses different constraints compared to the pure pitch tracking task, since in our scenario the pitch and its harmonics are shifted by an arbitrary frequency, requiring an open range search for all possible shifts. 

The contributions of this paper are two-fold. First, we introduce and discuss our new approach to carrier frequency difference estimation named ''Rake''. It is based on accumulated log-energy values and 
enables the classification on significantly shorter segments of speech compared to existing approaches, e.g., \cite{Ganapathy2013}. Second, an  efficient implementation in the power cepstrum domain is proposed to reduce the computational demands of the approach. Finally, in the experiments we evaluate the proposed algorithm on real \gls{SSB}  \gls{HF} recordings and also compare it to a state-of-the-art pitch tracking algorithm and a frequency difference estimation algorithm.

The paper is organized as follows: In Sec.~\ref{SEC:Rake} our features for carrier frequency difference estimation are derived, followed by Sec.~\ref{SEC:Implementation} where we discuss details of the implementation in the power cepstrum domain. In Sec.~\ref{SEC:PitchEx} and Sec.~\ref{SEC:Experiments} the experimental results are discussed. The paper ends by drawing some conclusions in Sec.~\ref{SEC:Conclusion}.


\section{Rake Approach} \label{SEC:Rake}
We are given a demodulated \gls{SSB} \gls{HF} signal, of which we assume that it has already been pre-processed by a speech activity detection unit, e.g., by the DNN-based approach from \cite{Heitkaemper2020}, such that only segments with active speakers are regarded in the following.
Note, that these segments consist of voiced and unvoiced speech, as well as short pauses. In the spectral domain the pitch and its harmonics are clearly visible in a spectrogram, if the \gls{STOI} value \cite{Taal2011STOI} is above $0.5$. However, many \gls{SSB} transmissions have much worse \gls{STOI} values and the pitch contours are occluded by noise or even completely erased, requiring noise robust algorithms and approaches (visit \cite{Heitkaemper2020a} for example signals).

In the following we propose  to estimate the carrier frequency difference by a method that is based on locating the pitch and its harmonics  in the noisy speech spectrogram. To this end it uses a filterbank with adjustable and time-varying center frequencies, that correspond to the fundamental frequency and its harmonics. One can view the filtering operation as a  rake that is  pulled in time direction through the logarithmic \gls{PSD} values $\log\{|X(t,f)|^2\}$ of the signal's \gls{STFT} $X(t,f)$, where $t$ denotes the frame index and $f$ the frequency bin index to collect the energy at the pitch frequency and its harmonics. The relevant frequency bin indices at the $t$-th frame for a hypothetical carrier frequency difference $\fD$,  pitch $f_P(t)$  and the corresponding pitch harmonics are given by $\fD + \tau \cdot f_{P}(t)$, where $(\tau \in [1, \tau_{\text{max}}])$.   To account for the limited frequency resolution of the \gls{STFT} analysis, not only the frequency bin itself but also a small range around it is considered by introducing the frequency deviation parameter $\nu$. So, the logarithmic \gls{PSD} values of the pitch including the harmonics are given by 
\begin{align} \label{EQ:PSDTerms}
&\Psi_{\nu}^\tau(t,{f}_{P}(t),\fD) = \log\{|X(t,\fD + \tau \cdot f_{P}(t)+\nu)|^2\}.
\end{align}
These values are weighted by factors $\omega(\tau,\nu)$, which depend on the harmonic index $\tau$ and the distance $\nu$ to the filter center, and are summed by
\begin{align} \label{EQ:GammaSum}
& \Gamma(t,f_{P}(t),\fD)  = \sum_{\tau=1}^{\tau_{\text{max}}} \sum_{\nu=-W}^{+W} \omega(\tau,\nu) \cdot \Psi_{\nu}^\tau(t,f_{P}(t),\fD).
\end{align}

For each frequency difference $\fD$ a different sequence of pitch values $\bm{f}_P = [f_{P}(0),\ldots f_{P}(T-1)]$ is optimal in the sense that the summation of $\Gamma(t,f_{P}(t),\fD)$ along  $t$ reaches a maximum. Stated differently, the maximization of \eqref{EQ:GammaSum} will yield an estimate of $\fD$. This is achieved by the following three steps. First, for each time instance $t$, the maximum across the possible pitch hypotheses $f_{P}(t) \in [f_{P,\min},f_{P,\max}]$ is computed:
\begin{align} \label{EQ:Max}
f'_{P}(t, \fD) = \underset{f_{P}(t)}{\argmax} \left\{ \Gamma(t,f_{P}(t),\fD)\right\} \\ 
\Gamma'(t,\fD) = \Gamma(t,f'_{P}(t, \fD),\fD).
\end{align}
$\Gamma'(t,\fD)$ is the maximum log-energy value for a given demodulation shift hypotheses $\fD$, and the corresponding pitch hypothesis is $f'_{P}(t, \fD)$. Here, $f_{P,\min}$ and  $f_{P,\max}$ denote the frequency bins corresponding to the assumed minimum (\SI{50}{Hz}) and maximum (\SI{400}{Hz}) pitch frequencies.

Next, a summation over time results in the accumulated log-energy per carrier frequency difference hypothesis $\fD$: 
\begin{align} \label{EQ:accumulatedLogEnergy}
& \widehat{\Gamma}(\fD)  = \sum_{t=0}^{T-1}  \Gamma'\left(t,\fD \right).
\end{align}
Assuming a single speaker scenario, the maximum of $\widehat{\Gamma}(\fD)$ is selected as the most likely hypotheses for the demodulation shift $\widehat{f}_{D}$ with 
\begin{align} \label{EQ:Argmax}
\widehat{f}_{D} = \underset{f_{D}\in \Omega_{\fD}}{\argmax} \left\{ \widehat{\Gamma}(\fD)\right\} 
\end{align}
with $\Omega_{\fD}$ denoting the set of candidate frequency differences,
and the corresponding  pitch hypotheses sequence is given by
\begin{align}
\label{EQ:PitchHyp}
\widehat{\bm{f}}_{P} = [f'_{P}(0, \widehat{f}_{D}),\ldots, f'_{P}(T-1, \widehat{f}_{D})].
\end{align}
As reported in several publications, e.g., in \cite{Zahorian08}, and also observed in our own experimental recordings, some audio segments have only a very weak or even no pitch at all, although the harmonics are clearly observable. To take account of this observation the weight of the pitch $\omega(0,\nu)$ is only half of $\omega(1,\nu)$, i.e., the weight of the first harmonic. Furthermore, all other harmonics are weighted with $\omega(\tau,\nu) \propto \frac{1}{\tau}$, whereby all filters are designed in a triangular shape.

The summation in \eqref{EQ:GammaSum} gives a similar pitch detection feature as the \gls{SHC} used in the YAAPT algorithm \cite{Zahorian08}, whereby \eqref{EQ:GammaSum} is defined as a sum of logarithmic \gls{PSD} values and \gls{SHC} is a sum over a product of magnitude spectral values. The log-spectral domain formulation causes a dynamic range reduction and improves the numerical stability. 

Since \eqref{EQ:GammaSum} extends the pitch tracking problem towards an open range search by introducing the unknown parameter $\fD$, the computational complexity of the problem is increased by a factor proportional to the size $|\Omega_{\fD}|$ of the set of candidate values for the frequency difference. Hence, reducing the computational complexity becomes an important task which in our case is handled by interpreting \eqref{EQ:GammaSum} in the cepstral domain as shown in the next section.


\section{Implementation} \label{SEC:Implementation}
The computationally expensive evaluation of the terms in \eqref{EQ:GammaSum} can be interpreted as a correlation of the logarithmic \gls{PSD} values $\log\{|X(t,f)|^2\}$ with a filter function
\begin{align} \label{EQ:FilterBank}
	h(f,f_P) = \sum_{\tau=1}^{\tau_{\text{max}}} \sum_{\nu=-W}^{W} \omega(\tau,\nu) \cdot \gamma(f-\tau \cdot f_P(t)-\nu),
\end{align}
along the frequency axis, where $\gamma(.)$ denotes the unit impulse. This interpretation is similar to the harmonic sieves proposed in \cite{Gerlach2014}. The correlation, which has to be carried out for all $f_P(t) \in [f_{P,\min}, f_{P,\max}]$, can be efficiently computed by applying a \gls{FFT}, i.e., by  moving to the  power cepstral domain, and using  the  Overlap-Save-Method. 

\begin{figure}[htb]
	\centering \footnotesize
	\def\svgwidth{.95\columnwidth}
	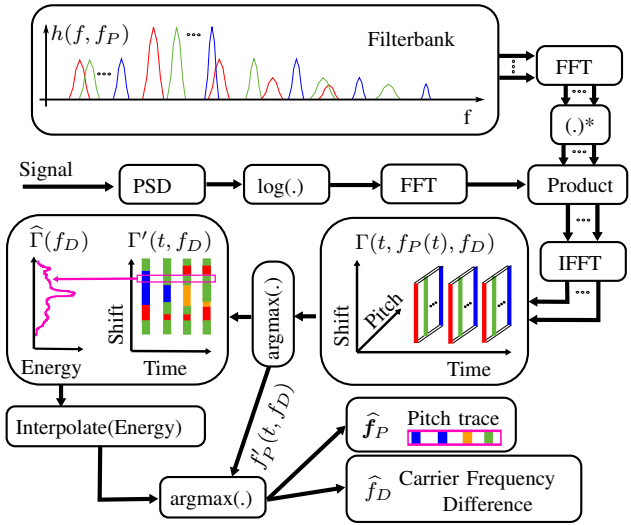
	\caption{Rake-PC: Block diagram showing power cepstral domain correlation, carrier frequency difference and pitch trace estimation.}
	\label{Fig:rakesimplified}
\end{figure}

Fig.~\ref{Fig:rakesimplified} shows a block diagram of the overall algorithm. This implementation is denoted as ''Rake-PC'' (Rake Power Cepstrum) in the following.
The upper part of the figure illustrates the realization of the correlation of the log-PSD with the set of filters, where each filter correponds to an assumed value of $f_P$, in the PSD domain. The resulting $\Gamma(t,f_{P}(t),\fD)$ depends on time frame $t$, the assumed pitch frequency $f_P(t)$, and the carrier frequency shift $\fD$. Next, the optimal value of the pitch is determined according to Eq.~\eqref{EQ:Max}, followed by a summation along the time axis, Eq.~\eqref{EQ:accumulatedLogEnergy}. The resolution of the resulting $\widehat{\Gamma}(\fD)$ is limited by the \gls{STFT} size, as the shift is given by its bin index. This can be overcome by interpolating the accumulated log-energy terms, e.g., by  spline interpolation. The subsequent argmax operation from \eqref{EQ:Argmax} yields the final estimate $\hat{f}_D$, whose resolution is no longer limited by the FFT size. 

Note, that the first maximum operation as stated in \eqref{EQ:Max}, is carried out independently for each time frame $t$, a clear shortcoming of the method, as it does not account for the inertia of the vocal cords, that results in smooth pitch trajectories. Introducing a-priori knowledge to account for the lowpass characteristics of the pitch trajectory, e.g., by using a simple first order Markov chain as proposed in \cite{Gerkman2010}, would improve the pitch tracking precision, however, at the cost of a significantly increased computational complexity.

The values of $\fD$ are quantized by the FFT resolution and the maximum
search from \eqref{EQ:Argmax} is restricted to the frequency bins that belong to the frequency range \SI{0}{Hz} to \SI{3500}{Hz} for a signal sampled at $\SI{8}{kHz}$.
The upper limit is motivated by the fact that a speech signal requires approximately a $\SI{500}{Hz}$ bandwidth to be intelligible and that the regarded harmonics have to fit in the considered frequency range. Signals with negative offsets $\fD$ are not considered because they are characterized by a significant signal loss of the lower frequencies and remain unintelligible without signal reconstruction approaches.

The availability of pitch trace estimates $f'_P(t,f_D)$, $t=0,\ldots , T-1$, offers the opportunity to discard maxima in $\widehat{\Gamma}(f_D)$ that are caused by narrow-band digital transmissions instead of speech. As human speech is characterized by a time-variant pitch, while digital transmissions operate with a fixed frequency, the two can be discerned by the variance of  $f'_P(t,f_D)$. If it falls below a threshold, the detected signal is probably not speech and consequently discarded.


\section{Pitch tracking experiments} \label{SEC:PitchEx}
\begin{figure}[b]
	\centering
	\input{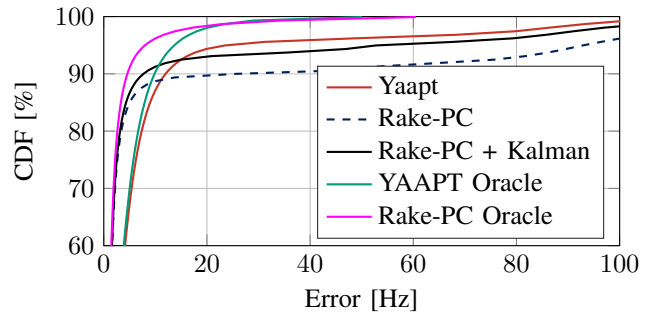}
	\caption{Cumulative density function (CDF) of pitch estimation error per frame on PTDB-TUG database \cite{PTDB-TUG2011}.}
	\label{fig:wizzard}
\end{figure}
To evaluate the proposed approach w.r.t.\ its capabilities of tracking the human pitch we used the PTDB-TUG database from \cite{PTDB-TUG2011} and compared our approach to the YAAPT algorithm implementation  \cite{Zahorian08}. The results are given in Fig.~\ref{fig:wizzard}. The Rake-PC pitch tracker achieves in more than 85\% of all pitch containing speech segments a higher precision than the YAAPT algorithm. However, in the remaining 15\% of the segments the performance stays way below YAAPT. This difference can be attributed to the sophisticated post-processing of YAAPT (multi pitch candidate selection process, non-linearities to restore missing pitches, application of temporal restrictions) which is missing in Rake-PC. If a Kalman filter is applied to the pitch trajectory estimated by Rake-PC, the performances difference can be compensated for to a great degree.

Fig.~\ref{fig:wizzard} also shows the results of an oracle experiment, where it was  allowed to multiply the pitch tracking results by a factor of $2$ or $0.5$ to compensate for mistakenly selecting a harmonic or subharmonic as the pitch frequency. Both algorithms benefit from the oracle, with Rake-PC achieving higher gains and even outperforming YAAPT. From this control experiment it can be concluded that the majority of the large errors in pitch estimation are caused by a wrong classification of pitch harmonics and sub-harmonics to be the pitch. This is a typical error to be handled by post-processing. But this misinterpretation has no impact on the task of carrier frequency difference estimation, because we are only interested in the sum of the PSD values at pitch and pitch harmonics. 


\section{Experiments on HAM radio data} \label{SEC:Experiments}

\begin{figure}[t]		
	\input{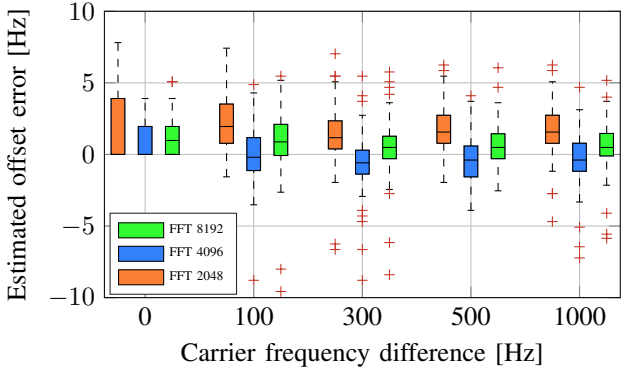}	
	\caption{Difference between estimated and ground truth carrier frequency difference for three FFT sizes. Length of speech activity was ${\geq}\SI{10}{s}$. No crosstalkers present.} 	\label{fig:cmp_FFT}	
\end{figure}


We have set up a transmission system between our amateur radio station in Paderborn and several other distant ham radio stations across Europe, transmitting utterances from the LibriSpeech corpus \cite{libri20}. Kiwi-\gls{SDR} devices \cite{kiwi20} at the distant  stations were utilized to demodulate the received \gls{SSB} \gls{HF} signals and  send the recorded audio signals back to our servers via a websocket connection. Audio markers had been added to the signal to allow for an automated time alignment between the transmitted and received signals, easing the annotation and segmentation of the data \cite{Heitkaemper2020a}.

For the transmissions a beacon, callsign DB0UPB, was used, which was supervised by a human to avoid  interference with other ham radio stations. The \gls{HF} signals are \gls{SSB} modulated using the \gls{LSB} with a  bandwidth of $\SI{2.7}{kHz}$ at carrier frequencies of $\SI{7.06}{MHz}-\SI{7.063}{MHz}$ and $\SI{3.6}{MHz} -\SI{3.62}{MHz}$. To simulate a carrier frequency difference the demodulation frequency of the transmitter and the receiver were selected to differ by values from the set $\fD = [0,100,300,500,1000]$. 
Although the original speech samples have a sampling rate of \SI{16}{kHz}, and the Kiwi-\gls{SDR} samples the data at \SI{12.001}{Hz}, the finally emitted data is band-limited to $\SI{2.7}{kHz}$ (\gls{ITU} regulations) which introduces a loss of the upper frequencies in case of \gls{LSB} transmission depending on the carrier frequency difference $\fD$. 
The data set has a total size of 23:31 hours of which 3:28 hours contain speech activity.

\subsection{Carrier Frequency Difference Estimation}
In Fig.~\ref{fig:cmp_FFT} the error between the estimated difference $\widehat{f}_{D}$ and the ground truth difference is depicted. For this experiment the length of the speech segments was between $\SI{10}{s}$ and $\SI{27}{s}$.  The system works reliable with errors below $\pm \SI{5}{Hz}$, which is an error that is not perceivable by humans \cite{Clark2013}.

\begin{figure}[t]
	\centering
	\resizebox{\columnwidth}{!}{
%
%
\definecolor{mycolor1}{rgb}{0.00000,0.44700,0.74100}%
\definecolor{mycolor2}{rgb}{0.85000,0.32500,0.09800}%
\definecolor{mycolor3}{rgb}{0.92900,0.69400,0.12500}%
\definecolor{mycolor4}{rgb}{0.49400,0.18400,0.55600}%
\definecolor{mycolor5}{rgb}{0.46600,0.67400,0.18800}%
\begin{tikzpicture}

\begin{axis}[%
width=1.7in,
height=1.6in,
at={(0in,0in)},
scale only axis,
bar width=0.2in,
area legend,
xmin=0.5,
xmax=5.5,
xtick={1,2,3,4,5},
xticklabels={{< 0.5},{0.5-1},{1-2},{2-10},{> 10}},
xlabel={Speech length [s]},
xmajorgrids,
ymin=0,
ymax=100.5,
ylabel={Error class affiliation [\%]},
ylabel style={yshift=-0.1cm},
ymajorgrids,
title style={font=\bfseries, yshift=-0.2cm},
title={Harmonic/Spectral Envelop},
legend style={legend cell align=left,align=left,draw=white!15!black}
]
\addplot[ybar stacked,draw=black,fill=mycolor1] plot table[row sep=crcr] {%
1	12.5290023201856\\
2	13.4963768115942\\
3	26.3403263403263\\
4	22.2619047619048\\
5	35.343618513324\\
};
\addplot[ybar stacked,draw=black,fill=mycolor2] plot table[row sep=crcr] {%
1	5.10440835266821\\
2	11.231884057971\\
3	10.8558108558109\\
4	11.1904761904762\\
5	15.2875175315568\\
};
\addplot[ybar stacked,draw=black,fill=mycolor3] plot table[row sep=crcr] {%
1	14.15313225058\\
2	24.6376811594203\\
3	22.5441225441225\\
4	22.7777777777778\\
5	29.8737727910238\\
};
\addplot[ybar stacked,draw=black,fill=mycolor4] plot table[row sep=crcr] {%
1	9.51276102088167\\
2	13.2246376811594\\
3	9.89010989010989\\
4	11.1507936507937\\
5	8.69565217391304\\
};
\addplot[ybar stacked,draw=black,fill=mycolor5] plot table[row sep=crcr] {%
1	58.7006960556845\\
2	37.4094202898551\\
3	30.3696303696304\\
4	32.6190476190476\\
5	10.7994389901823\\
};
\end{axis}

\begin{axis}[%
width=1.7in,
height=1.6in,
at={(2.0in,0.02471in)},
scale only axis,
bar width=0.2in,
area legend,
xmin=0.5,
xmax=5.5,
xtick={1,2,3,4,5},
xticklabels={{< 0.5},{0.5-1},{1-2},{2-10},{> 10}},
xlabel={Speech length [s]},
xmajorgrids,
ymin=0,
ymax=100.5,
ymajorgrids,
title style={font=\bfseries, yshift=-0.2cm},
title={Rake-PC},
legend style={at={(0.97,0.03)},anchor=south east,legend cell align=left,align=left,draw=white!15!black}
]

\addplot[ybar stacked,draw=black,fill=mycolor1] plot table[row sep=crcr] {%
1	40.3712296983759\\
2	60.2355072463768\\
3	76.5567765567766\\
4	87.7777777777778\\
5	96.0729312762973\\
};
\addlegendentry{\footnotesize \textless~5 Hz};

\addplot[ybar stacked,draw=black,fill=mycolor2] plot table[row sep=crcr] {%
1	13.9211136890951\\
2	14.6739130434783\\
3	10.0899100899101\\
4	4.40476190476191\\
5	0.981767180925666\\
};
\addlegendentry{\footnotesize 5 -- 10 Hz};

\addplot[ybar stacked,draw=black,fill=mycolor3] plot table[row sep=crcr] {%
1	15.0812064965197\\
2	7.15579710144928\\
3	2.76390276390276\\
4	0.555555555555556\\
5	0.140252454417952\\
};
\addlegendentry{\footnotesize 10 -- 50 Hz};

\addplot[ybar stacked,draw=black,fill=mycolor4] plot table[row sep=crcr] {%
1	5.80046403712297\\
2	2.17391304347826\\
3	0.532800532800533\\
4	0.158730158730159\\
5	0\\
};
\addlegendentry{\footnotesize 50 -- 100 Hz};

\addplot[ybar stacked,draw=black,fill=mycolor5] plot table[row sep=crcr] {%
1	24.8259860788863\\
2	15.7608695652174\\
3	10.0566100566101\\
4	7.1031746031746\\
5	2.80504908835905\\
};
\addlegendentry{\footnotesize \textgreater~100 Hz};
\end{axis}
\end{tikzpicture}%
}
	\caption{Dependency of error class affiliation on speech segment length. Implementation of Harmonic/Spectral Envelop following \cite{Ganapathy2013}. FFT size set to $4096$.}
	\label{fig:ErrorClassAffiliation}
\end{figure}
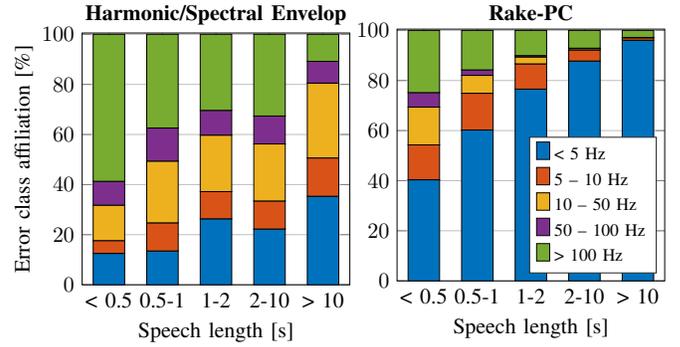

For shorter speech segments the error increases as can be seen in Fig.~\ref{fig:ErrorClassAffiliation}, where we grouped the estimation errors in five classes. In that figure we compared our approach to the harmonic/spectral envelope approach of \cite{Ganapathy2013}, which we implemented on our own since no open source implementation was available.
It can be observed that the proposed approach achieves lower estimation errors, both on short and longer speech utterances.

\begin{figure}[b]
	\centering
	\input{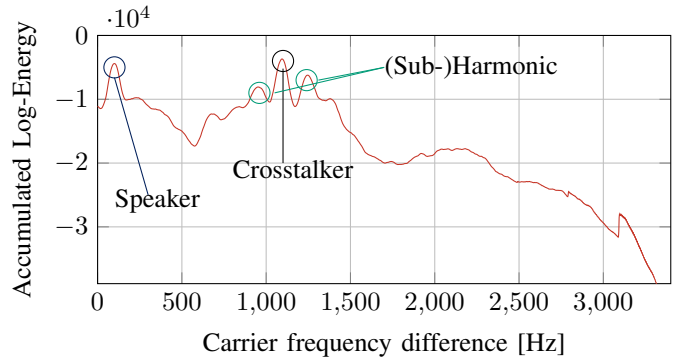}
	\caption{Accumulated logarithmic energy in case of two parallel speakers. Crosstalker is visible as secondary maximum and its (sub-)harmonics.}
	\label{fig:multipleparallelspeakerpsdtrack}	
\end{figure}

\subsection{Parallel speakers and harmonic errors}
A small amount of our recordings include the special case of a concurrent speaker at a higher frequency. This is a challenging, however likely scenario to be encountered in practice.
Fig.~\ref{fig:multipleparallelspeakerpsdtrack} depicts the accumulated logarithmic energy $\widehat{\Gamma}(\fD)$, Eq.~\eqref{EQ:accumulatedLogEnergy}, versus candidate frequency difference $\fD$.  Both, the speaker at $\SI{100}{Hz}$ and the interfering crosstalker at $\SI{1098}{Hz}$ are visible as maxima in the accumulated log-energy. Also two secondary maxima are visible next to the crosstalker which we attribute to harmonics/subharmonics and a possible  non-linearity in the transmission system. So extending the Rake-PC towards concurrent speaker tracking and identification seems to be possible, similar to multi-speaker tracking in diarization \cite{Hogg2019} or localization tasks \cite{Gerlach2014}.

\subsection{Processing time}
\begin{figure}[htb]
	\centering
%
%
\definecolor{mycolor1}{rgb}{1.00000,0.00000,1.00000}%
\begin{tikzpicture}

\begin{axis}[%
width=1.4in,
height=1.5in,
at={(0in,0in)},
scale only axis,
xmin=0.8,
xmax=3.2,
xtick={1,2,3},
xticklabels={{2048},{4096},{8192}},
xlabel={FFT},
xmajorgrids,
ymode=log,
ymin=0.01,
ymax=100,
yminorticks=true,
ylabel={Real time factor},
ymajorgrids,
yminorgrids,
legend style={at={(1.1,0.225661)},anchor=south west,legend cell align=left,align=left,draw=white!15!black}
]
\addplot [color=red,dashed,mark=*,mark options={solid,fill=red,draw=red}]
  table[row sep=crcr]{%
1	2.8\\
2	11.02\\
3	44.28\\
};
\addlegendentry{\footnotesize Rake (Single Core)};

\addplot [color=orange,dashed,mark=*,mark options={solid,fill=orange,draw=orange}]
  table[row sep=crcr]{%
1	0.076\\
2	0.2716\\
3	0.95\\
};
\addlegendentry{\footnotesize Rake-PC (Single Core)};

\addplot [color=green,dashed,mark=*,mark options={solid,fill=green,draw=green}]
  table[row sep=crcr]{%
1	0.02\\
2	0.0582\\
3	0.2326\\
};
\addlegendentry{\footnotesize Rake-PC (Multi Core)};

\addplot [color=mycolor1,dashed,mark=diamond*,mark options={solid,fill=mycolor1,draw=mycolor1}]
  table[row sep=crcr]{%
1	0.04\\
2	0.045\\
3	0.055\\
};
\addlegendentry{\footnotesize Harmonic/Spectral \cite{Ganapathy2013}};

\addplot [color=green,only marks,mark=diamond*,mark options={solid,fill=black,draw=black}]
  table[row sep=crcr]{%
3	0.074549638525412\\
};
\addlegendentry{\footnotesize YAAPT \cite{Zahorian08}};

\end{axis}
\end{tikzpicture}%
	\caption{Real time factors for different approaches and \gls{FFT} sizes, including single and multi core implementation for a block shift of \SI{20}{ms}. (AMD Ryzen 5 3600, 6-Core, 32GB RAM, Matlab)}
	\label{fig:realtime}	
\end{figure}
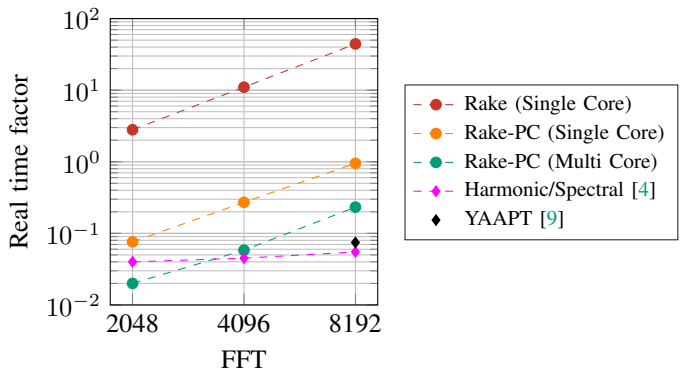

In Fig.~\ref{fig:realtime} the real time factors of the proposed approach, our implementation of \cite{Ganapathy2013} and the reference implementation of the YAAPT algorithm from \cite{Zahorian08} are given. The implementation following Fig.~\ref{Fig:rakesimplified} (''Rake-PC (Single Core)'') improves the real time factor significantly compared to the direct implementation (''Rake (Single Core)''). The overall processing time can be further reduced by a  straight forward parallel implementation (''Rake-PC (Multi Core)''). For a \gls{FFT} size of 2048  Rake-PC has a similar real time factor as \cite{Ganapathy2013} and \cite{Zahorian08}.

\section{Conclusions} \label{SEC:Conclusion}
In this paper we have presented an approach that estimates frequency differences between the mixing oscillator in the HF transmitter, that moves the baseband signal to the carrier frequency, and the oscillator in the receiver that converts the radio signal back to baseband. It is based on  tracking the pitch and its harmonics in the received baseband speech signal in an open range search method. Experiments on real data from \gls{HF} transmissions show promising performance, both in terms of precision and computational complexity.


\balance
\renewcommand*{\bibfont}{\small}
\bibliographystyle{IEEEtran}
\bibliography{eulib}

\end{document}